\documentclass[aps]{revtex4}
\usepackage{amssymb}
\usepackage{amsmath}
\usepackage{epsfig}

\numberwithin{equation}{section}
\newcommand{\rot}{\nabla\!\times\!}

\begin{document}

\title{Scattering of slow-light gap solitons with charges
 in a two-level medium}
\author {J. Leon$^1$, P. Anghel-Vasilescu$^1$, F. Ginovart$^2$, N. Allegra$^1$}
\affiliation{$^1$ {\em Laboratoire de Physique Th\'eorique et Astroparticules},
Universit\'e de Montpellier 2 (CNRS-IN2P3) Montpellier (France)\\
$^2$ {\em Laboratoire FOTON}, Universit\'e de Rennes,
ENSSAT (CNRS) F22305 Lannion (France)}

To appear in \textit{J. Phys. A: Math. Theor.}

\begin{abstract}The Maxwell-Bloch system describes a quantum two-level medium
interacting with a classical electromagnetic  field by mediation of the the
population density. This population density variation is a purely quantum
effect which is actually at the very origin of nonlinearity. The resulting
nonlinear coupling possesses particularly interesting consequences at the
resonance (when the frequency of the excitation is close to the transition
frequency of the two-level medium) as e.g. \textit{slow-light gap
solitons} that result from the nonlinear instability of the evanescent wave at
the boundary. As nonlinearity couples the different  polarizations of the
electromagnetic field, the slow-light gap soliton is shown to experience
effective scattering whith \textit{charges} in the medium, allowing it for
instance to be trapped or reflected. This scattering process is understood
qualitatively as being governed by a nonlinear Schr\"odinger model in an
external potential related to the charges (the electrostatic permanent
background component of the field). \end{abstract}

\maketitle

\section{Introduction}

In the field or interaction of radiation with matter, the resonant scattering of
an electromagnetic pulse with a structure imprinted in a medium (as a periodic
grating) has become widely studied for its rich and novel nonlinear
mechanisms\cite{Yeh1}. Such studies have been developped in nonlinear optics
where the medium is a periodic arrangement of dielectrics (Bragg medium) and
where the nonlinearity results from Kerr effect\cite{Agrawal}, or in
Bose-Einstein condensates where the periodic structure is produced by external
applied field (thus called an optical lattice)\cite{Morsh,Meacher}, or else in
photorefractive media where an externally applied electric potential modifies
locally the index value\cite{Yeh2}.

In a Bragg medium, the underlying model is the Maxwell equation for the
electromagnetic field where the optical index varies along the propagation
direction and where the constitutive equations (relating field and polarization)
take into account the third order suceptibility. A simplified model is obtained
then in the slowly varying envelope approximation: the nonlinear Schr\"odinger
equation embedded in a potential which represents the structure of the medium.
In a Bose-Einstein condensate, the model is the Gross-Pitaevskii equation which
is again a nonlinear Schr\"odinger equation in an external potential
representing the optically generated structure (the optical grating). Last, in
a photorefractive medium, the variations of the applied  electrostatic field
induce a local variation of the optical index, and possibly also of the group
velocity dispersion, which eventually results again in a NLS like model with
variable coefficients. 

It is remarkable that these situations share the same model and that it is
precisely the simplest nonlinear version of the paradigm model for wave
scattering, the Schr\"odinger equation of quantum mechanics, namely
\begin{equation}\label{NLS-pot}
 i\frac{\partial \psi}{\partial t}+
\frac{\partial^2 \psi}{\partial z^2}+|\psi|^2\psi-V\,\psi=0
\end{equation}
where the wave function $\psi=\psi(z,t)$ is submitted both to self-interaction
(nonlinearity) and to scattering with the external potential $V$ (depending on
space $z$ and possibly also on time $t$). 

In fact the common original mechanism is the coupling of an electromagnetic
radiation of a given frequency $\omega$, to a quantum two-level system whose
transition frequency $\Omega$ is close to $\omega$. The interaction is then
mediated by the variation of the population densities of the levels which makes
it a nonlinear process. The semi-classical model of such a process is the
Maxwell-Bloch (MB) system which can be written \cite{Fain,Allen, Kaplan,Pantell}
\begin{align}
& \frac{\partial^2{\bf P}}{\partial t^2}+\Omega^2{\bf P}=
-\left[ \frac{2}{\hbar}\Omega\Lambda|\vec\mu_{12}|^2\right] \, 
 N\,{\bf E} ,\label{MB1}\\
& \hbar\Omega\frac{\partial N}{\partial t}=
2\Lambda{\bf E}\cdot\frac{\partial{\bf P}}{\partial t},
\label{MB2}\\
& \rot\rot {\bf E}+\frac{\eta^2}{c^2}
\frac {\partial^2{\bf E}}{\partial t^2} =
-\mu_0\Lambda\frac {\partial^2{\bf P}}{\partial t^2}.\label{MB3}
\end{align}
in the case when the dipolar momentum $\vec \mu_{12}$ is assumed to be parallel
to the applied electric field $\bf E$ (at least in average). The constant
$\Lambda=(\eta^2+2)/3$ is the Lorentz local field correction,
$\eta$ is the optical index of the medium and $\Omega$ the transition frequency.
The dynamical variables are the real valued 3-vectors $\bf E$ (electric
field) and $\bf P$ (polarization source), and the inversion of population
density $N$. The above first two equations are a convenient rewriting of the
Bloch equation of quantum mechanics for the density matrix $\rho$ where one
defines
\begin{equation}
{\bf P}=N_0(\rho_{21}\vec \mu_{12}+\rho_{12}\vec \mu_{21})
,\quad N=N_0(\rho_{22}-\rho_{11}),
\end{equation}
with $N_0$ being the density of state of active elements (note that hermiticity
guaranties that $N$ and ${\bf P}$ are real valued). To be complete, the vector
$\vec\mu_{12}$ is defined from the two eigenstates $|u_i\rangle$ of the
unperturbed
Hamiltonian $H_0$ by
\begin{equation}
 \vec\mu_{12}=\left\langle u_1|e\vec r|u_2\right\rangle,\quad
H_0u_i={\cal E}_iu_i,\quad  {\cal E}_2-{\cal E}_1=\hbar\Omega,
\end{equation}
when the electric dipolar Hamiltonian is $H_0-e\vec r\cdot\vec E$.

In the above MB system, the variations of the dynamical variable $N$ (inversion
of population density) is on the one side a purely quantum effect and on the
other side the very mechanism of nonlinearity. Indeed, the linear limit (small
field values) comes from assuming $N$ constant (in fact $N=-N_0$ if the medium
is in the ground state), and MB reduces to the Lorentz classical linear
model (equation (\ref{lorentz}) below). Moreover, the structure of equation
(\ref{MB2}) implies effective coupling of all 3 components of the field, a
property which will allow us to couple a polarized tranverse electromagnetic
component to the longitudinal electrostatic one.

Inclusion of a particular physical situation consists for the MB system in a
convenient choice of the set of initial and boundary values for a particular
structure of the fields (as e.g. a propagation in one given direction). The
freedom in such choices makes the MB system a very rich model that has not
finished to offer interesting results. Many different limits of MB have been
studied, for instance the case of a weak coupling (under-dense media) allows for
{\em self-induced transparency} (SIT) of a light pulse whose peak frequency is
tuned to the resonant value $\Omega$, when a linear theory would predict total
absorption \cite{machan}\cite{sit-rep}. The limit model in the slowly varying
envelope approximation results to be integrable \cite{lamb} and  to have the
{\em mathematical} property of transparency: any fired pulse having an area
above a threshold evolves to a soliton, plus an asymptotically vanishing
background \cite{abkane}. Description of SIT process within the inverse spectral
transform is fruitful here because the incident pulse, which is physically
represented by a boundary datum, maps to an {\em initial value problem} on the
infinite time line. 

Away from the resonance and for dense media, a widely used approach models the
dynamical nonlinear properties of pulse propagation by selecting propagation in
one direction. The resulting {\em reduced Maxwell-Bloch system} has again the
nice property of being integrable \cite{eilbeck,sasha} (also when detuning and
permanent dipole are included \cite{moloney}). As a consequence the properties
of the {\em gap soliton} such as pulse reshaping, pulse slowing, pulse-pulse
interactions,  are fairly well understood, more especially as the reduced MB
system possesses explicit N-soliton solutions \cite{hynne}. Others interesting
features include pulse velocity selection \cite{branis}. However, unlike the SIT
model, the reduced MB system happens to be integrable for an {\em initial pulse
profile} which makes it almost useless to study the scattering of an
\textit{incident} light pulse.

To describe explicit scattering process, the reduced MB equation was
replaced by the {\em coupled-mode Maxwell-Bloch} system where the electric field
envelope contains both right-going and left-going slowly varying components
\cite{mantsyzov}.  Although the adequation of the model to the physical
situation is a very difficult question often left apart, the approach
allowed to prove of existence of gap 2$\pi$-pulses in the presence of
inhomogeneous broadening, and to discover {\em ``optical zoomerons''}. Moreover
numerical simulations have demonstrated  the possibility of {\em ``storage of
ultrashort optical pulses''} \cite{xiao}. This storage can moreover be
externally managed to release the stored pulse and thus create a {\em ``gap
soliton memory''} with a two-level medium \cite{melnikov}. 

The coupled-mode approach has been also applied to understand the properties of
{\em resonantly absorbing Bragg reflectors} introduced in \cite{kozekin} and
further studied in \cite{malomed,sjohn}, which consist in periodic arrays of
dielectric films separated by layers of a two-level medium. Very recently, the
coupled-mode MB system has been used to model {\em ``plasmonic Bragg gratings''}
\cite{Raether} in nanocomposite materials where a dielectric is imbedded in a
periodic structure of thin films made of metallic nanoparticules \cite{ildar}.

Beyond such a set of approximations, the Maxwell-Bloch system offers a natural
intrinsic nonlinear coupling of the various field components. It has been
demonstrated for instance that, for a uni-directional propagation, the coupling
of two transverse electromagnetic components can be described in the slowly
varying envelope approximation by a (non-integrable) set of two coupled
nonlinear Schr\"odinger equations \cite{gino}. Thus the question of the analytic
expression of the fundamental soliton solution is still open. However such a
coupling process allows to conceive a method to manipulate light pulses with
light by propagating a pulse on a background made of a stationary wave, when
wave and pulse are orthogonally polarized \cite{ramaz-prl}.

The problem of the generation of a pulse living in the stop gap of the two-level
medium has been solved in \cite{gap-sol-pra} by exciting the medium at one
end with a cavity standing wave. As the frequency is in the stop gap, the
result is an evanescent wave, at least at the linear level. It happens that the
evanescent wave is nonlinearly unstable, which results in the \textit{nonlinear
supratransmission} process \cite{nst-prl} that generates gap solitons
propagating in the two-level medium at a fraction of light velocity.

We demonstrate here that the effective coupling of a tranverse electromagnetic
(polarized) field to a longitudinal electrostatic component having a
permanent background (representing local charges) allows to scatter a slow
light gap soliton (SLGS) and to trap it in the medium or to make it move
backward. This result is obtained by first studying the MB system in a
particular situation of a propagation in a given direction (say $z$) of a field
which is linearly polarized in the tranverse direction (say $x$) and which
interacts, through the variations of the population density, with a
longitudinal component. Then by appropriate boundary values for the tranverse
component and initial data for the longitudinal one, we perform
numerical simulations that show the propagation and scattering of the SLGS. 

We thus demonstrate that the presence of charges in a definite region of space
(obtained for example by an applied electrostatic potential or with a doped
semiconductor, or else with inclusion of metallic nanoparticles) produces a
dynamical interaction with an electromagnetic radiation by means of a nonlinear
coupling through the plasma wave field component (spontaneously generated out of
the permanent electrostatic background). It is worth mentionning that this work
makes use of the results of \cite{gap-sol-pra} where a SLGS is generated in
homogeneous two-level media and that the perturbative asymptotic analysis
follows \cite{gino} and will not be detailed (see the appendix). Last, another
way of manipulating light pulses have been proposed in \cite{ramaz-prl} by
creating in the two-level medium a standing electromanetic field background
(inside the passing band) which is a completely different physical process.

The scattering is finally given a simple meaning by writing a nonlinear
Schr\"odinger model for the envelope of the electromagnetic field in an external
\textit{potential} resulting from the electrostatic permanent background.
Interestingly enough, the usual asymptotic perturbative analysis of the system
leads to a deformed nonlinear Schr\"odinger system which allows to calculate the
correct expression of the \textit{initial vacuum}. However the model does not
furnish an accurate description of the dynamical properties of the SLGS. This is
the result of the intrinsic nature of the method which separates the orders of
perturbation when the original model naturally couples them.

\section{The model}
\subsection{Dimensionless form.}

The system (\ref{MB1}-\ref{MB3}) can be written under a dimensionless form by
defining first the new space-time variables
\begin{equation}\label{dimens-spatial}
 \vec r\,'=\frac{\eta}{c}\Omega\,\vec r,\quad t'=\Omega t 
\end{equation}
and the new field variables
\begin{equation}\label{dimens-field}
{\bf E}'=\sqrt{\frac{\epsilon}{ W_0}}\,{\bf E},\quad
{\bf P}'=\Lambda\sqrt{\frac{1}{\epsilon W_0}}\,{\bf P},\quad
N'=\frac{N}{N_0},\quad 
\vec\mu_{12}\,'=\frac{\vec\mu_{12}}{\mu}
\end{equation}
where by definition of the optical index $\epsilon=\epsilon_0\eta^2$, and where
\begin{equation}
 W_0=N_0\hbar\Omega/2,\quad
\mu^2=\vec\mu_{12}^*\cdot\vec\mu_{12}
\end{equation}
Note that $W_0$ is a reference energy density. Then the MB system
 eventually reads (forgetting the \textit{primes})
\begin{align}
& \frac{\partial^2{\bf P}}{\partial t^2}+{\bf P}=
-\alpha N\,{\bf E} ,\nonumber\\
&\frac{\partial N}{\partial t}={\bf E}\cdot
\frac{\partial{\bf P}}{\partial t},\label{MB}\\
& \rot\rot {\bf E}+\frac {\partial^2{\bf E}}{\partial t^2} =
-\frac {\partial^2{\bf P}}{\partial t^2}.\nonumber
\end{align}
We check that the unique remaining coupling constant $\alpha$, defined by
\begin{equation}\label{alpha}
 \alpha=\frac{2N_0}{\hbar\Omega}\,\frac{\mu^2}{\epsilon}\,\Lambda^2.
\end{equation}
is indeed dimensionless. The length $\mu$ of the dipolar moment is in units of
$C\cdot m$, the energy $\hbar\Omega$ in $J$ while the permittivity $\epsilon$
can be expressed in $C^2\cdot J^{-1}\cdot m^{-1}$. Using finally that $N_0$ is a
density in $m^{-3}$ effectively leads to a dimensionless $\alpha$. The actual
dynamical variables are  the dimensionless quantities ${\bf E}(\vec r,t)$, 
${\bf P}(\vec r,t)$ and $N(\vec r,t)$, namely 7 scalar real variables, fully
determined by the MB system (\ref{MB}) with convenient initial boundary value
data. Note that the scaled inversion of population density $N$ now varies from
$N=-1$ (fundamental) to $N=1$ (excited).

\subsection{Polarized waves.}

\begin{figure}[ht]
 \centerline{\epsfig{file=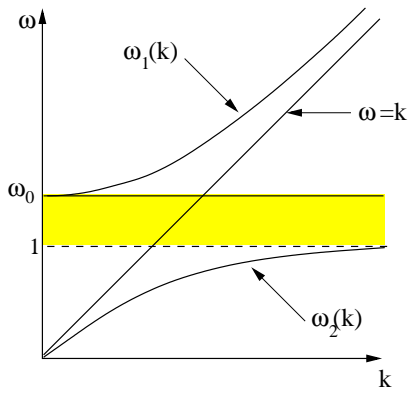,width=0.3\linewidth}} 
\caption { Plot of the linear dispersion relation of Maxwell-Bloch for
$\alpha=1$. The two curves are the two solutions $\omega_1(k)$ and $\omega_2(k)$
of expression (\ref{disp}). We have also plotted the plasma frequency $\omega_0$
and the free photon line $\omega=k$.   Solitons will be generated in the
forbidden band gap  $]1,\,\omega_0]$.}
\label{fig:disp}
\end{figure}

We restrict our study to unidirectional propagation for a field which possesses
a linearly polarized electromagnetic component $E(z,t)$ and a longitudinal
component  $F(z,t)$. Namely we assume the following structure
\begin{equation}\label{structure}
 {\bf E}=\left( \begin{array}{c} E(z,t)\\ 0\\ F(z,t)
\end{array}\right)\quad\Rightarrow\quad
{\bf P}=\left( \begin{array}{c} P(z,t)\\ 0\\ Q(z,t)
\end{array}\right),
\end{equation}
for which the system (\ref{MB}) reads
\begin{align}
&  P_{tt}+P=\alpha(1-n)E,\quad & Q_{tt}+Q=\alpha(1-n)F,\nonumber\\
& E_{tt}-E_{zz}=-P_{tt} ,\quad & F_{tt}=-Q_{tt},\label{MB-ES1}\\
&n_t=EP_t+FQ_t.\nonumber
\end{align}
where we have defined the scaled density of excited states
\begin{equation}\label{def-n}
 n(z,t)=1+N(z,t),
\end{equation}
The compatibility of the chosen structure with the MB model can be
demonstrated at the linear level obtained for $N=-1$ (or $n=0$)
\begin{equation}\label{lorentz}
\frac{\partial^2{\bf P}}{\partial t^2}+{\bf P}=\alpha\,{\bf E} ,\quad
\rot\rot {\bf E}+\frac {\partial^2{\bf E}}{\partial t^2} =
-\frac {\partial^2{\bf P}}{\partial t^2}.
\end{equation}
which is nothing but the Lorentz model. The general fundamental solution 
possessing the structure (\ref{structure}) can be written
\begin{align}
& {\bf E}=
\left(\begin{array}{c} E_0\\0\\0\end{array}\right)e^{i(\omega t-kz)}+
\left( \begin{array}{c} 0\\ 0\\ f(z)\end{array}\right)e^{i\omega_0 t}+
\left( \begin{array}{c} 0\\ 0\\ F_0(z)\end{array}\right),
\label{linear_sol}\\
& {\bf P}=\frac{\omega_0^2-1}{1-\omega^2}
\left(\begin{array}{c} E_0\\0\\0\end{array}\right)e^{i(\omega t-kz)}+
\left( \begin{array}{c}0\\0\\ -f(z)\end{array}\right)e^{i\omega_0 t}+
\left( \begin{array}{c} 0\\ 0\\\alpha F_0(z)\end{array}\right).\nonumber
\end{align}
The above fields are constituted of 3 basic parts: the first one is the
tranverse electromagnetic component where $k$ is given by the dispersion
relation (plotted on fig.\ref{fig:disp})
\begin{equation}
k^2=\omega^2\,\frac{\omega^2-\omega_0^2}{\omega^2-1},\quad
\omega_0^2=1+\alpha,\label{disp}
\end{equation}
the second one is the plasma wave of frequency $\omega_0$ and arbitrary
$z$-dependent amplitude $f(z)$, the third one is the electrostatic field
$F_0(z)$ that represents the permanent charges. Indeed the Gaus theorem relates
$F_0$ to the (dimensionless) charge density $\rho$ by 
$\nabla\cdot({\bf E}+{\bf P})=\rho=(1+\alpha)\partial_zF_0(z)$.

\subsection{Initial data.}

The above structure of the linear solution shows that it is possible to
eliminate the variable $Q(z,t)$ by integrating one equation in the system
(\ref{MB-ES1}). Actually the equation $F_{tt}=-Q_{tt}$  results from the
original equation $\rot{\bf H}=\partial_t{\bf D}$ (no current of charges). As
the propagation is along $z$, the third component of the curl vanishes,
therefore $(\partial_t{\bf D})_z=0=\partial_t(F+Q)$, which allows to express
$Q(z,t)$ as $Q(z,t)=-F(z,t)+C(z)$, where the integration constant $C(z)$ remains
to be computed. It is naturally determined from the initial conditions. We
consider here a situation where the presence of stationary charges produces a
permanent electrostatic component $F_0(z)$, namely when
\begin{equation}\label{F-phi}
 F(z,t)=\phi(z,t)+F_0(z),\quad \phi(z,0)=0.
\end{equation}
Then the structure (\ref{linear_sol}) of the linear solution then implies 
the following expression for the longitudinal component of
the polarization
\begin{equation}\label{Q}
 Q(z,t)=\alpha F_0(z)-\phi(z,t),
\end{equation}
where, in the absence of initial electromagnetic irradiation, we have the
initial conditions $\{n(z,0)=0,\, \phi(z,0)=0\}$. We now demonstrate that the
initial \textit{vacuum} $\{E=P=0,n=0,\phi=0\}$ and $\{F=F_0(z),Q=\alpha
F_0(z)\}$ is \textit{nonlinearly stable} as soon as $F_0$ is below some
threshold.

\subsection{Nonlinear stability.}

The stability of the \textit{initial vacuum} is proved for the dynamical
variable $\phi(z,t)$ without applied external electromagnetic field $E(z,t)$
(the $x$-component). Replacing expressions (\ref{F-phi}) and (\ref{Q}) in the
system (\ref{MB-ES1}), the equation for the electrostatic wave $\phi(z,t)$
becomes
\begin{equation}
 \phi_{tt}+\omega_0^2\phi=\alpha n(\phi+F_0).
\end{equation}
Without applied electromagnetic wave ($E=P=0$), the equation for $n(z,t)$ (the
density of population of the excited state) can be integrated taking into
account $n(z,0)=0$ and $\phi(z,0)=0$ to give
\begin{equation}
 n=-\frac{1}{2}\phi^2-\phi F_0,
\end{equation}
which, in the evolution for $\phi$, provides the closed equation
\begin{equation}
 \phi_{tt}+(\omega_0^2+\alpha F_0^2)\phi+\frac{\alpha}{2}\phi^2 (\phi+3F_0)=0.
\end{equation}
This is an equation for an anharmonic oscillator whose frequency depends on the
\textit{parameter} $F_0(z)$. The solution corresponding to
$\phi(z,0)=0$ and $\phi_t(z,0)=0$ is obviously $\phi(z,t)=0$, which shows that
the initial conditions are consistent. To study the stability, we rewrite the
above oscillator as
\begin{equation*}
 \frac{d}{dt}\left[\frac{1}{2}\phi_t^2+{\cal E}_p\right]=0,\quad
{\cal E}_p=\frac{1}{2}(\omega_0^2+\alpha F_0^2)\phi^2+
\frac{\alpha}{2}F_0\,\phi^3+\frac{\alpha}{8}\,\phi^4.
\end{equation*}
The condition of \textit{nonlinear stability} is reached when
the function ${\cal E}_p(\phi)$ has one single minimum in
$\phi=0$ (initial condition). The derivative 
\begin{equation*}
 \frac{d{\cal E}_p}{d\phi}=\frac{\alpha}{2}\, \phi\left[
\phi^2+3F_0\phi+2(\frac{\omega_0^2}{\alpha}+F_0^2)\right] .
\end{equation*}
vanishes always at $\phi=0$, and there will be no other solutions as soon as 
$\max_z\{F_0^2\}<8(1+\alpha)/\alpha$ (in our case $\alpha=1$, i.e.
$8(1+\alpha)/\alpha=16$). With a single root at $\phi=0$, the potential 
${\cal E}_p(\phi)$ has a single minimum in $\phi=0$. However, when 
$\max_z\{|F_0|\}$ exceeds the threshold value (namely the value 4 with
$\alpha=1$), the potential ${\cal E}_p(\phi)$ becomes a double well
with a \textit{local maximum} at point $\phi=0$ which is thus unstable (and
might be of interest to study). Thus here we shall consider values of $F_0$ such
that $\max_z\{|F_0|\}<4$ such that nonlinear stability condition be always
satisfied.

Thus we have obtained the initial state reached by the medium submitted to
electrical potential (charges), namely to a permanent longitudinal component
$F_0(z)$ which has been shown to generate a permanent polarization $\alpha
F_0(z)$ accordingly with (\ref{Q}). These data constitute then the initial
condition (initial equilibrium ground state) that is now inserted in the model
itself.

\subsection{Final model.}

\begin{figure}[ht]
 \centerline{\epsfig{file=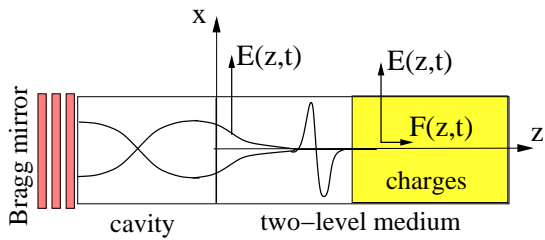,width=0.4\linewidth}} 
\caption { Principle of a situation where the tranverse generated
electromagnetic gap soliton would scatter with a permanent longitudinal
background $F_0(z)$ created by a region with charges (doped smiconductor,
applied static electric potential, inclusion of metallic nanoparticles...) where
a longitudinal component $F(z,t)$ is generated by nonlinear coupling.}
\label{fig:device}
\end{figure}

The  model studied from now on eventually reads
\begin{align}
&  P_{tt}+P=\alpha(1-n)E,\quad & \phi_{tt}+\omega_0^2\phi=
\alpha n(\phi+F_0),\nonumber\\
& E_{tt}-E_{zz}=-P_{tt} ,\quad & n_t=EP_t-(\phi+F_0)\phi_t,\label{MB-final}
\end{align}
and it is completed by the following initial values (ground state at rest)
\begin{align}
& E(z,0)=0,\quad E_t(z,0)=0,\quad P(z,0)=0,\quad P_t(z,0)=0,\nonumber\\
& \phi(z,0)=0,\quad \phi_t(z,0)=0,\quad n(z,0)=0.\label{init}
\end{align}
Two boundary values for the electromagnetic components are now needed, chosen
following \cite{gap-sol-pra} as representing an electromagnetic cavity coupled
to the two-level medium by using its stop gap as one side \textit{mirror} (the
cavity works at frequency $\omega$ in the gap).  We then assume an open end
$z=L$ (vanishing of the magnetic component), therefore we set
\begin{equation}\label{bound-E}
 E(0,t)=a\,\sin(\omega t),\quad E_z(L,t)=0,
\end{equation}
with $\omega\in[1,\,\omega_0]$. Last, $F_0(z)$ is a given function that results
from the presence of charges, (created e.g. with a doped smiconductor, an
applied static electric potential, inclusion of metallic nanoparticles,...), and
that needs to be evaluated in a given physical context. We present in
fig.\ref{fig:device} the scheme of principle of the device that corresponds to
our problem.

\section{Numerical simulations}

Our purpose is simply here to show that a given function $F_0(z)$ acts as a
scattering potential on the dynamics of the \textit{slow light gap soliton}. To
that end one needs first to generate the SLGS. It is done by applying the
principle of nonlinear supratransmission \cite{nst-prl} to the MB system as
described in \cite{gap-sol-pra}. There it is proved that the boundary value
(\ref{bound-E}) at a frequency in the stop gap, which would linearly produce an
evanescent wave in the medium, is actually \textit{unstable} above
a threshold $a_s$ given by
\begin{equation}\label{threshold}
a_s=4\sqrt{2\omega_0\,\frac{\omega_0-\omega}{\omega_0^2+3}}
\end{equation}
where $\omega$ is the driving frequency in the stop gap $[1,\,\omega_0]$.
\begin{figure}[ht]
 \centerline{\epsfig{file=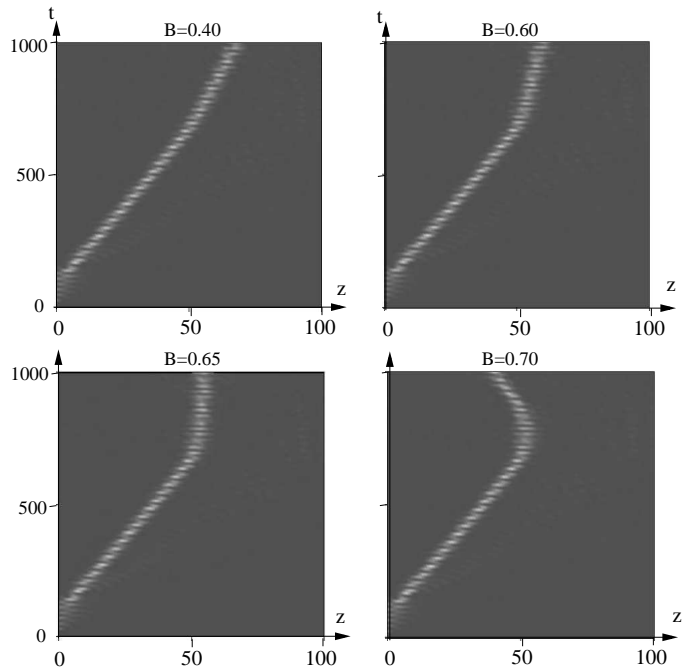,width=0.5\linewidth}} 
\caption {Scattering of the SLGS generated by the boundary  (\ref{bound-E})
with $a=0.4$ and $\omega=1.4$ onto the step function $F_0(z)$ of
(\ref{pot-step}) with $z_0=50$ and values of step height $B$ as indicated.
These are intensity plots of the energy density flux $-EP/2$ where the
brightest regions have value $0.3$.} \label{fig:scattMB}
\end{figure}

Let us simply mention that the threshold of nonlinear supratransmission is
obtained from the nonlinear Schr\"odinger limit of MB by assuming a driving
frequency close to the gap edge $\omega_0$. For instance it is shown in
\cite{gap-sol-pra} this threshold is correct to a high precision close to the
edge (typically for $\omega=1.4$ with the edge $\omega_0=\sqrt{2}$), precision
that decreases to $15\%$ at the frequency $1.2$. But the very question is not
the value of the threshold but its \textit{existence}, source of gap soliton
generation. Note that one needs to ensure compatibility of boundary values
(\ref{bound-E}) with initial data (\ref{init}) and thus we always start the
boundary driving smoothly at $t>0$ such that $E(0,0)=0$. Note also the the
stability of the static solution is used to determine a stable \textit{vacuum}
solution inside the region with charges, which is disconnected from the
instability of the evanescent wave living in the region without charges.

We display in fig.\ref{fig:scattMB} the result of the scattering of a SLGS onto
a step potential
\begin{equation}\label{pot-step}
 F_0(z)=B\,\theta(z-z_0),
\end{equation}
where $\theta$ is the Heaviside function. We have used as boundary
(\ref{bound-E}) the following driving (at frequency $\omega=1.4$)
\begin{equation}\label{pulse-bound}
E(0,t)=0.2[\tanh(0.2(t-20))-\tanh(0.5(t-150))]\cos(\omega t),
\end{equation}
The scattering mechanism will be analyzed in the next section, let us simply
mention here that the SLGS is very robust and has been checked to (almost)
maintain its amplitude and frequency across the step, the only varying parameter
being its velocity. This property will allow us to understand the existence of
a threshold step height above which the SLGS is reflected. Below this threshold
the SLGS simply slows down to ajust to a different medium.

\section{A phenomenological model}

\subsection{The NLS model}
\begin{figure}[ht] \centerline
{\epsfig{file=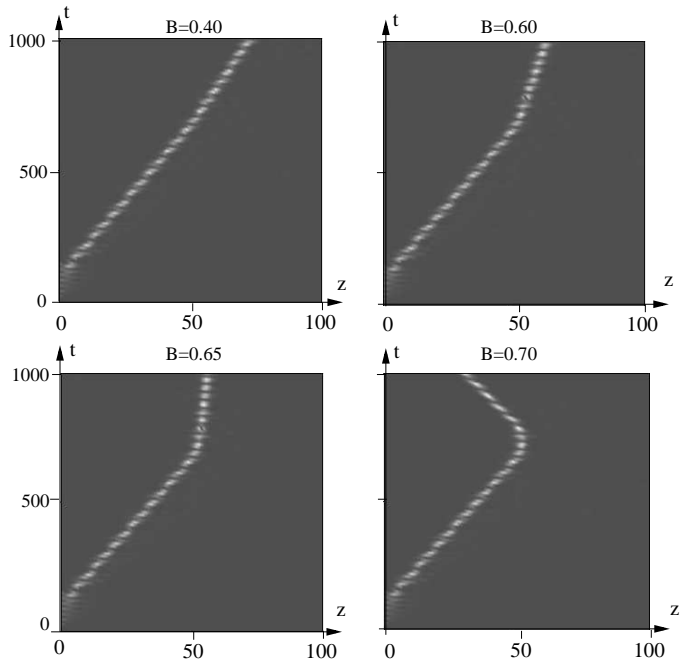,width=0.5\linewidth}} 
\caption {\it Test of the soliton scattering by use of the NLS model
(\ref{NLS-pheno}) with boundary driving (\ref{boundNLS}) where $a=0.4$ and
$\omega=1.4$, and with $V(z)$ given by (\ref{V-pheno}) where
$F_0(z)=B\,\theta(z-40)$.  We have plotted the energy flux density $-EP/2$ given
here by (\ref{EP_nls}), which concur with simulations of Maxwell-Bloch of
fig.\ref{fig:scattMB}.}
\label{fig:scattNLS}
\end{figure}

To describe the gap soliton scattering in a simple and efficient way, we propose
here the following model
\begin{equation}\label{NLS-pheno}
-2i\omega_0\psi_t+\frac{\omega_0^2-1}{\omega_0^2}\, \psi_{zz}+
\frac{1}{2}(\omega_0^2+3)|\psi|^2\psi=V(z)\,\psi.
\end{equation}
similar to (\ref{NLS-pot}) and obtained from (\ref{NLS}) by replacing the action
of $\varphi$ by that of an external potential $V(z)$. The related
initial-boundary value problem is naturally deduced from relation
(\ref{E-asymp}) and boundary value (\ref{bound-E}) as
\begin{equation}\label{boundNLS}
 \psi(z,0)=0,\quad \psi(0,t)=-i\frac{a}{2}e^{i(\omega-\omega_0)t}.
\end{equation}
This constitutes a \textit{phenomenological} model where value of the applied
potential $V(z)$ is obtained by comparing numerical simulations of the above
equation with those of the Maxwell-Bloch system (\ref{MB-final}). We have
obtained that
\begin{equation}\label{V-pheno}
V(z)=\frac{1}{2}\, \frac{(\omega_0^2-1)^2}{\omega_0^2+4}  F_0(z)^2,
\end{equation}
provides a quite accurate description of the scattering process. The above
particular form has been inspired by the nature of the asymptotic expansion
in eq.(\ref{psit}) with the structure (\ref{nut}), especially for the dependence
in $F_0^2$, but the precise expression has been worked out to generate
scattering processes similar to those obtained from Maxwell-Bloch.  This is
illustrated by fig.\ref{fig:scattNLS} obtained by solving (\ref{NLS-pheno})
with the above potential and boundary value. Then to compare to the plots of
fig.\ref{fig:scattMB}, we contruct the electromagnetic energy density flux
$-EP/2$ which from (\ref{E-asymp}) is given by
\begin{equation}\label{EP_nls}
 -\frac{1}{2}EP\simeq \frac{1}{2} \left(
\psi\,e^{i\omega_0t}+\bar\psi\,e^{-i\omega_0t}\right) ^2.
\end{equation}

\subsection{Scattering simple rules}

The interest of having such a simplified model is to derive explicit formulae
for the scattered gap soliton. The NLS equation in a potential has been widely
studied, e.g. by use of the inverse spectral theory \cite{balak}, and especially
when the potential is a Dirac delta function representing a local inhomogeneity
\cite{newell}\cite{boris}\cite{holmer}, which has also been done for discrete
systems \cite{peyrard}. An interesting application is the study, with a
stochastic series of delta functions, of the competition between disorder and
nonlinearity \cite{kivshar}\cite{Flach}. In our case the step (or barrier)
height is not
small compared to soliton amplitude ($V_0$ is of the order of $max|\psi|^2$)
and one cannot call to perturbative approach to get analytical description of
the scattering process. However, we shall obtain by quite elementary arguments,
an expression of the scattered soliton velocity in terms of the incident
velocity and the parameters of the medium.

We are interested in potentials $V(z)$  piecewise constant
(in regions sufficiently larger than the soliton extension) for which the NLS
equation (\ref{NLS-pheno}) possesses approximate soliton solution in each
region, far enough from the discontinuities. The soliton solution of
(\ref{NLS-pheno}) for any $V(z)=V_0$ constant is given by
\begin{equation}
 \psi_s(z,t)=\frac{Ae^{i(\beta z-\mu t)}}{\cosh[\gamma(z-vt)]}
\end{equation}
with the following 3 relations linking the 5 coefficients $A$ (amplitude),
$\beta$ (wave number), $\mu$ (frequency), $\gamma$ (stiffness) and $v$
(velocity). These are
\begin{align}
\beta=-\frac{\omega_0^3}{\omega_0^2-1}\, v,\label{beta_sol}\\
\gamma^2=\frac{A^2}{4}(\omega_0^2+3)\frac{\omega_0^2}{\omega_0^2-1},
\label{gamma_sol}\\
V_0=(\gamma^2-\beta ^2)\frac{\omega_0^2-1}{\omega_0^2}-2\mu\omega_0.
\label{V_sol}
\end{align}
One important consequence of the above relations is obtained by elimination of
$\beta$ and $\gamma$ and writes
\begin{equation}\label{v_sol}
 v^2\frac{\omega_0^4}{\omega_0^2-1}=\frac{A^2}{4}
(\omega_0^2+3)-2\omega_0\mu-V_0,
\end{equation}
which furnishes the soliton velocity from its amplitude $A$ and carrier
frequency $\mu$. Note that the solution is a \textit{gap} soliton which means 
$\mu>0$. 

Now we assume, accordingly with observations of a number of numerical
simulations, that the amplitude $A$ and the frequency $\mu$  do not change
across a step from a region 1 where $V=0$ to a region 2 where $V=V_0$. The
soliton is produced in region 1 by the boundary driving and are
measured for each numerical experiment as in \cite{gap-sol-pra} (it is actually
easier to measure $v$ and $\mu$ and to deduce $A$). Then the scattering process
is understood in a quite simple manner. First formula (\ref{v_sol}) immediately
implies the existence of a particular threshold value of the constant $V_0$ for
which the velocity in region 2 would vanish and above which this soliton
is not a solution. This threshold, called $V_{th}$, is thus obtained by
setting $v=0$ in (\ref{v_sol}) and reads
\begin{equation}\label{Vth1}
 V_{th}=\frac{A^2}{4}(\omega_0^2+3)-2\omega_0\mu.
\end{equation}
To be clear, this threshold indicates that the NLS equation (\ref{NLS-pheno})
with $V_0>V_{th}$ does not support a soliton solution with the given parameters
$A$ and $\mu$. So we expect a \textit{reflection} on the step of height $V_0$
as soon as $V_0>V_{th}$. 

Below the threshold, the velocity $v'$ in region 2 is deduced from
(\ref{v_sol}) and (\ref{Vth1}) as
\begin{equation}\label{vitesse}
V_0<V_{th}\quad : \quad {v'}^2=v^2-V_0\frac{\omega_0^2-1}{\omega_0^4},
\quad  V_{th}=v^2\, \frac{\omega_0^4}{\omega_0^2-1}.
\end{equation}
Here the threshold value $V_{th}$ given in (\ref{Vth1}) is
written in terms of the emitted SLGS velocity $v$ only by computing
hereabove the value of $V_0$ for which $v'=0$.

\subsection{Step potential}

In the case of the one-step potential (\ref{pot-step}) we have
\begin{equation}\label{step}
z>z_0\,:\quad V_0=\frac{1}{2}\, \frac{(\omega_0^2-1)^2}{\omega_0^2+4}
\,B^2,
\end{equation}
and the velocity $v'$ reads from (\ref{vitesse})
\begin{equation}\label{vit2}
{v'}^2=v^2-\frac{B^2}{2\omega_0^4}\,\frac{(\omega_0^2-1)^3}{\omega_0^2+4}.
\end{equation} 
As a consequence  the threshold amplitude $B_{th}$ of the
electrostatic permanent background $F_0(z)=B\theta(z-z_0)$ can be written
\begin{equation}\label{Bth}
B_{th}=v\,\omega_0^2\sqrt{2}\,\frac{(\omega_0^2+4)^{1/2}}{(\omega_0^2-1)^{3/2}}.
\end{equation}
\begin{figure}[ht] \centerline
{\epsfig{file=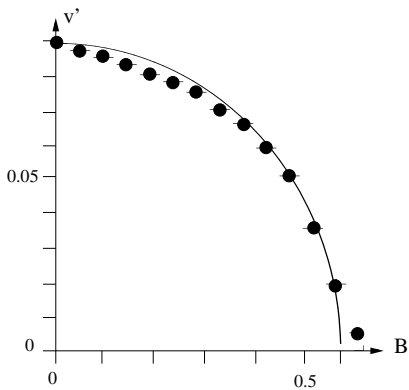,width=0.3\linewidth}} 
\caption {\it Soliton velocity $v'$ beyond the step location $z_0$ in terms of
the step height $B$ according to formula (\ref{vit2}) plotted as the
continuous curve, and measured from simulations of Maxwell-Bloch
(points).} \label{fig:vit}
\end{figure}

We have tested these formulae against numerical simulations of the Maxwell-Bloch
system (not simply the NLS model), which is summarized in figure \ref{fig:vit}.
Clearly the small discrepancy between soliton scattering and elementary formula
(\ref{vit2}) comes on the one side from the assumptions of conservation of shape
across the step, on the other side from losses by phonon emission. Note that
formula (\ref{vit2}) relates the kinetic energy of the soliton in the two media
through the potential energy $V_0$ of the obstacle. Another interesting aspect
of this formula is the fact that the scattered velocity $v'$ depends only on the
incident velocity $v$, once the step height $V_0$ and the coupling constant
$\alpha=\omega_0^2-1$ have been fixed.

Such an expression can be confronted to numerical simulations of Maxwell-Bloch
for different values of the coupling parameter $\alpha$ and different incident
SLGS velocities $v$ (obtained by varying the driving amplitude).
\begin{equation}\label{table}
\begin{array}{|c|c|c|c|c|c|}
\hline
\alpha & \omega &  a   &  v   & B_{th} &   B   \\
\hline
  1    & 1.40   & 0.40 & 0.088 & 0.61   & 0.65  \\
\hline
  1    & 1.40   & 0.42 & 0.111 & 0.77   & 0.72  \\
\hline
  1    & 1.40   &  0.44 & 0.133 & 0.92   & 0.90  \\
\hline
  1    & 1.40   & 0.46 & 0.154 & 1.07   & 1.00  \\
\hline
  2    & 1.70   & 0.60 & 0.175 & 0.69   & 0.70  \\
\hline 
  3    & 1.95   & 0.70 & 0.14 & 0.45   & 0.50  \\
\hline
  4    & 2.18   & 0.72 & 0.15 & 0.40   & 0.42  \\
\hline
  5    & 2.40   & 0.70 & 0.22 & 0.53   & 0.54  \\
\hline
 10    & 3.25   & 0.70 & 0.18 & 0.34   & 0.38  \\
\hline
\end{array}
\end{equation}
Some results are presented in table \ref{table} where $\alpha$ is the coupling
parameter, $a$ and $\omega$ are the chosen driving amplitude and frequency for
Maxwell-Bloch as defined in (\ref{bound-E}). Then $v$ is the measured soliton
velocity and $B_{th}$ the theoretical threshold for soliton reflection as given
by (\ref{Bth}). The effective threshold $B$ is then obtained by multiple
simulations of Maxwell-Bloch where $F_0(z)=B\theta(z-z_0)$ varying the value of
$B$. Such a result shows that the empirical formula (\ref{V-pheno}) for the
potential $V(z)$ entering the nonlinear Schr\"odinger model correctly describes
the scattering process within the chosen parameter range.

\section*{Conclusion and comments}

The main result we wish to emphasize is the demonstration that a slow light gap
soliton in a two-level medium can be manipulated by use of a permanent
electrostatic background field. Such a property may have interesting
applications as it is quite generic and fundamental: a two-level medium
constitutes the basic model for any medium submitted to monochromtic radiation
close to one transition frequency.

One interesting open problem is the description of the interaction process
through the method of perturbative asymptotic expansion. We have indeed
discovered that it is the interaction of processes of different orders which is
fundamental in the coupling of the electromagnetic field with the electrostatic
component. We have not been able to include such interaction in the perturbative
asymptotic method as by nature it decouples the different orders. Although the
simple nonlinear Schr\"odinger model (\ref{NLS-pheno}) describes quite well the
SLGS scattering of Maxwell-Bloch, one would appreciate to \textit{obtain} it by
some limit procedure. Actually the NLS equation is a natural limit, it is the
expression of the external potential (\ref{V-pheno}) representing the
electrostatic permanent background that should be rigorously derived.

One may of course play with different types of potentials, as for instance a
barrier of height $B$ and width $D$. In that case at a given height, and a given
incident soliton velocity, we observe tunnelling when the width value is below a
threshold (e.g. we have obtained tunnelling with parameters of
fig.\ref{fig:scattMB} at height $B=0.8$ for a width $D<4.5$). This is an
interesting problem, obviously related with the delta-function case
\cite{boris}, that will be considered in future studies.

Note finally that the soliton is generated from the boundary driving
(\ref{bound-E}) by the \textit{nonlinear supratransmission} mechanism which
is an instability of the evanescent wave \cite{leon-instab}. As a consequence,
the soliton characteristics are quite sensitive to the driving parameters:
frequency, amplitude, and shape. The fig.\ref{fig:scattMB} has been obtained
for the particular boundary value (\ref{pulse-bound}) and the soliton is
generated at a velocity $v=0.88$. A different shape of that driving, keeping
same amplitude and frequency, may generate a SLGS with a slightly different
velocity. Note that the fundamental question of the relation between soliton
characteristics and driving parameters is also an open problem.

\appendix
\section{Perturbative asymptotic analysis}

The purpose of this appendix is to provide usefull expressions for the
asymptotic model derived from Maxwell-Bloch as in \cite{gino} but here with
non-vanishing permanent fields. Although computations are lengthly, they are
standard and we list here only the relevant results.

\paragraph{General expressions}
It is
actually  convenient to consider the full 3-vector case and to reduce it to
our particular interest at the very end. The dimensionless Maxwell-Bloch system
(\ref{MB}) reduced to a propagation in the $z$-direction becomes
\begin{align}
\partial^2_t {\bf P}+  {\bf P}-\alpha {\bf E}=-\alpha  n
 {\bf E},\quad 
\partial_t n =  {\bf E}\cdot\partial_t  {\bf P},\nonumber\\
\left(\partial^2_t- \sigma\partial^2_z\right)
  {\bf E}= -\partial^2_t  {\bf P},\label{eq:basic}
\end{align}
where the singular matrix $\sigma$ is defined by
\begin{equation}
 \sigma=\left(\begin{array}{ccc}1&0&0\\0&1&0\\0&0&0\end{array}\right).
\end{equation}
The reductive perturbative expansion is a method to extract information about
the variations at first order of a formal expansion of the fields in terms of
new slow variables, see e.g. the tutorial paper \cite{leblond} and the refrences
therein. We seek \textit{slowly varying envelope approximation} solutions (SVEA)
for a carrier wave at frequency $\omega$ very close to $\omega_0$ (and in the
gap: $\omega<\omega_0$), thus we assume the expansion
\begin{align}
 {\bf E}=\sum_{j=1}^{\infty}\epsilon^j
\sum_{m=-j}^{+j} {\bf E}_j^m(\xi,\tau)e^{im\omega_0t},\nonumber\\
 {\bf P}=\sum_{j=1}^{\infty}\epsilon^j
\sum_{m=-j}^{+j}  {\bf P}_j^m(\xi,\tau)e^{im\omega_0t},\label{eq:expansion}\\
 n=\sum_{j=1}^{\infty}\epsilon^j\sum_{m=-j}^{+j}
    N_j^m(\xi,\tau)e^{in\omega_0t},\nonumber
\end{align}
in which the reality condition implies
\begin{equation}  {\bf E}_j^{-m}=\bar  {\bf E}_j^m,\quad
 {\bf P}_j^{-m}=\bar {\bf P}_j^m,\quad N_j^{-m}=\bar N_j^m.
\label{eq:reality}\end{equation}
An important issue is to include in the expansion the description of the
permanent background $F_0(z)$. Then we must assume that  ${\bf E}$ and ${\bf P}$
do possess zero-frequency modes, namely that ${\bf E}_j^0\ne 0$ and $ {\bf
P}_j^0\ne 0$. We shall find that these quantities must obey some general
constraint but are not governed by an equation (this is the reason why in
general they are assumed to vanish) which will allow us to fix them as external
data (scattering potential). The {\em slow variables} $\xi$ and $\tau$ are here
\begin{equation} 
\xi=\epsilon z,\quad \tau=\epsilon^2 t.\end{equation}
Indeed, the natural space variable is usually $\xi=\epsilon(z-vt)$ where $v$ is
the group velocity at the frequency of the carrier. Here this frequency is
$\omega_0$ for which the group velocity vanishes, as implied by the dispersion
relation (\ref{disp}). Note that this property ensures that the
boundary-value problem in the physical space maps effectively to a boundary
value problem in the \textit{slow space}.

Inserting the infinite series expansion (\ref{eq:expansion}) into the
system (\ref{eq:basic}) we obtain, after careful (tedious) algebraic
computations,  the following closed form system at order $\epsilon^3$
(remember ${\bf E}_1^{-1}=\overline{{\bf E}_1^{1}}$)
\begin{align}
-2i\frac{\omega_0}{\alpha}\partial_{\tau}{\bf E}_{1}^{1}+
\frac{\sigma}{\omega_0^2}
\partial_\xi^2 {\bf E}_{1}^{1}=
-N_{2}^{0}{\bf E}_{1}^{1}
+({\bf E}_{1}^{1}\cdot{\bf E}_{1}^{0}){\bf E}_{1}^{0}
+\frac{1}{2}({\bf E}_{1}^{1}\cdot{\bf E}_{1}^{1}){\bf E}_{1}^{-1},
\nonumber\\
\partial_\tau N_2^0
=\partial_\tau\left(
\frac{\alpha}{2}{\bf E}_{1}^{0}\cdot{\bf E}_{1}^{0}
+\frac{\omega_0^2+1}{\omega_0^2-1}{\bf E}_{1}^{1}\cdot{\bf E}_{1}^{-1}\right),
\label{NLS-N20}
\end{align}
together with the following constraints on the backgroung field 
${\bf E}_{1}^{0}$
\begin{equation}\label{pot_elec_stat}
\left(\begin{array}{ccc}\partial_\xi^2 &0&0\\
0&\partial_\xi^2 &0\\0&0&\partial_\tau^2\end{array}\right)
{\bf E}_{1}^{0}=0.
\end{equation}
The principle of the method is then to express all relevant quantities in terms
of the solution of the nonlinear evolution (\ref{NLS-N20}). These are given by
\begin{align}
{\bf E}= &\epsilon\left( {\bf E}_{1}^{0}+{\bf E}_{1}^{1}e^{i\omega_0t}+
{\bf E}_{1}^{-1}e^{i\omega_0t}\right)+{\cal O}(\epsilon^2)
\label{E-exp}\\
{\bf P}= &\epsilon\left(\alpha{\bf E}_{1}^{0}-{\bf E}_{1}^{1}e^{i\omega_0t}-
{\bf E}_{1}^{-1}e^{-i\omega_0t}\right)+{\cal O}(\epsilon^2)
\label{P-exp}\\
 n= &\epsilon^2(N_{2}^{0}
-{\bf E}_{1}^{1}\cdot{\bf E}_{1}^{0}e^{i\omega_0t}
-\frac{1}{2}{\bf E}_{1}^{1}\cdot{\bf E}_{1}^{1}e^{2i\omega_0t}
-{\bf E}_{1}^{-1}\cdot{\bf E}_{1}^{0}e^{-i\omega_0t}\nonumber\\
&-\frac{1}{2}{\bf E}_{1}^{-1}\cdot{\bf E}_{-1}^{1}e^{-2i\omega_0t}
)+{\cal O}(\epsilon^3),\label{n-exp}
\end{align}
where ${\bf E}_1^1$ and $N_2^0$ are  given in equations (\ref{NLS-N20}). The
structure of the external datum ${\bf E}_1^0$  resulting from the constraint
(\ref{pot_elec_stat}) reads
\begin{equation}
 {\bf E}_{1}^{0}=\left(\begin{array}{c}
E_{1x}^{0}(\tau) \\E_{1y}^{0}(\tau)\\ E_{1z}^{0}(\xi)\end{array}\right).
\end{equation}
where the dependences can be arbitrarily fixed.

\paragraph{Electrostatic potential}

By inverse scaling we may come back to the original physical variables $z$ and
$t$, and we restrict our study to the case (\ref{structure}) where the
longitudinal component $F(z,t)=\phi(z,t)+F_0(z)$. 
Rewriting then expressions (\ref{E-exp})-(\ref{n-exp}) in physical variables, we
get at first order (namely $\epsilon$ for ${\bf E}$ and ${\bf P}$,
$\epsilon^2$ for $n$)
\begin{align}
{\bf E}=\left( \begin{array}{c} 0\\ 0\\ F_0(z)
\end{array}\right)+
\left[\left( \begin{array}{c} \psi(z,t)\\ 0\\
\varphi(z,t)\end{array}\right)e^{i\omega_0t}+c.c.\right],
\label{E-asymp}\\
{\bf P}=\left( \begin{array}{c} 0\\ 0\\ \alpha F_0(z)
\end{array}\right)-
\left[\left( \begin{array}{c} \psi(z,t)\\ 0\\
\varphi(z,t)\end{array}\right)e^{i\omega_0t}+c.c.\right] ,
\label{P-asymp}\\
 n=\nu -\left[ F_0\varphi\,e^{i\omega_0t}+
\frac{1}{2}(\psi^2+\varphi^2)\,e^{2i\omega_0t}+c.c.\right] 
\label{n-asymp}
\end{align}
where we rename $(E_1^1)_x=\psi$, $(E_1^1)_z=\varphi$ and $N_2^0=\nu$. The
corresponding evolutions (\ref{NLS-N20}) eventually read
\begin{align}
-2i\omega_0\psi_t+\frac{\omega_0^2-1}{\omega_0^2}\, \psi_{zz}
=-\alpha \nu\psi+\frac{\alpha}{2}(\psi^2+\varphi^2)\bar\psi,
\label{psit}\\
-2i\omega_0\varphi_t-\alpha(F_0)^2\varphi
=-\alpha \nu\varphi+\frac{\alpha}{2}(\psi^2+\varphi^2)\bar\varphi,
\label{phit}\\
\nu_t =\partial_t\left(\frac{\alpha}{2}(F_0)^2
+\frac{\omega_0^2+1}{\omega_0^2-1}(|\psi|^2+|\varphi|^2)\right).
\label{nut}
\end{align}
Thanks to the initial data (\ref{init}), namely here to
$ \psi(z,0)=0$, $\varphi(z,0)=0$ and  $\nu(z,0)=0$,
we may integrate now the evolution of $\nu$ to get
\begin{equation}\label{nu-bare}
 \nu=\frac{\omega_0^2+1}{\omega_0^2-1}(|\psi|^2+|\varphi|^2).
\end{equation}
Inserted in the equations for $\psi$ and $\varphi$ it gives the final NLS-like
system
\begin{align}
-2i\omega_0\psi_t+&\frac{\omega_0^2-1}{\omega_0^2}\, \psi_{zz}+
\frac{1}{2}(\omega_0^2+3)|\psi|^2\psi=\nonumber\\
&\frac{1}{2}(\omega_0^2-1)\varphi^2\bar\psi-(\omega_0^2+1)
|\varphi|^2\psi,\label{NLS}\\
-2i\omega_0\varphi_t-&\alpha(F_0)^2\varphi+
\frac{1}{2}(\omega_0^2+3)|\varphi|^2\varphi=\nonumber\\
&\frac{1}{2}(\omega_0^2-1)\psi^2\bar\varphi-(\omega_0^2+1)
|\psi|^2\varphi.\label{varphi}
\end{align} 
The equation (\ref{NLS}) is  a nonlinear Schr\"odinger equation for $\psi$
(envelope of the tranverse electromagnetic component) in some \textit{external
potential} created by  $\varphi$, the envelope of the \textit{plasma wave}. The
applied permanent background $F_0(z)$ does not act \textit{directly} on $\psi$
but through the plasma wave $\varphi$ according to the dynamical equation
(\ref{varphi}).

The above asymptotic analysis has allowed to demonstrate that the elimination
of the variable $Q(z,t)$ by equations (\ref{F-phi}) and (\ref{Q}), namely by
comes naturally as the first order solution. Then the initial condition
$\phi(z,0)=n(z,0)=0$ was shown to be stable in the absence of applied
electromagnetic transverse component. By means of  (\ref{E-asymp}), such a set
of initial data maps to $\varphi(z,0)=0$. The problem is that the evolution
(\ref{varphi}) has the unique solution $\varphi(z,t)=0$ for initial data
$\varphi(z,0)=0$ and $\varphi_t(z,0)=0$ (representing an initial equilibrium
state). Thus we are not able to use the system of NLS-like equations
(\ref{NLS})(\ref{varphi}) to describe the gap soliton scattering with purely
vanishing initial data. 

This comes from the fact that the scattering process in Maxwell-Bloch is
initiated at higher orders by nonlinear coupling. Indeed, considering the MB
system (\ref{MB-final}) with initial-boundary value problem
(\ref{init})(\ref{bound-E}), it is clear that the interaction of $E$
(electromagnetic component) with the permanent electrostatic background $F_0(z)$
works with the birth of the plasma wave $\phi$. And $\phi$ grows on an
initial vacuum by mediation of the normalized density $n$ of population of the
excited level, which is an effect of second order (see the time evolution of
$n(z,t)$). But the method of multiscale asymptotic series \textit{separates the
orders} as indeed, at each order $\epsilon^n$ an equation is obtained
(independent of $\epsilon$). We thus cannot expect the method to describe
correctly such a situation.

We have pursued the series to order $\epsilon^5$ to check that indeed no
\textit{source term} appear. As expected from the method, we got a linear system
for the next order expansion with the variable coefficients $\varphi$ and
$\psi$,  given from the preceding order. This confirms that the method does 
not allow feedback interaction of higher orders  to the fundamental one
$\{\psi,\varphi\}$.

\textbf{Aknowledgements.} This work has been done as part of the programm
GDR 3073 \textit{Nonlinear Photonics and Microstructured Media}.

\end{document}